\documentclass[aps,prc,twocolumn,showpacs,nofootinbib]{revtex4-1}
\usepackage{graphicx,amsmath,amssymb,bm,color,physics}
\usepackage{epstopdf}

\newcommand{\beq}{\begin{equation}}
\newcommand{\eeq}{\end{equation}}

\allowdisplaybreaks

\begin{document}

\title{Path-integral Monte Carlo study of particles obeying quantum mechanics and classical statistics}

\author{William G. Dawkins}
\author{Alexandros Gezerlis}
\affiliation{Department of Physics, University of Guelph, Guelph, Ontario N1G 2W1, Canada}

\begin{abstract}

Ultracold atomic systems have been of great research interest in the past, with more recent attention being paid to systems of mixed species. In this work we carry out non-perturbative Path Integral Monte Carlo (PIMC) simulations of $N$ distinguishable
particles at finite temperature, which can be thought of as an ultracold atomic system containing $N$ distinct species. We use the PIMC approach to calculate thermodynamic properties of particles interacting via hard-sphere and hard-cavity potentials. The first problem we study is a two-particle system interacting via a hard-sphere and hard-cavity interaction in order to test the effectiveness of two approximations for the thermal density matrix corresponding to these potentials. We then apply the PIMC method to a system of many hard-sphere particles under periodic boundary conditions at varying temperature in order to calculate the energy per particle, pressure, and specific heat of the system. We examine how finite-size effects impact the results of PIMC simulations of hard-sphere particles and when the thermodynamic limit has been reached. Our results provide microscopic benchmarks for a system containing distinguishable particles, which can be thought of as a limiting case for ultracold atomic systems of mixed species.   

\end{abstract}

\maketitle

\section{Introduction}

The study of cold atomic systems has been of great interest over the past century or so and has given great insight into fundamental quantum phenomena. More recently, experimentalists have been able to probe cold Fermi gases
and even tune the interaction between two atoms via the use of Feshbach resonances. This allowed unprecedented access 
to detailed features or novel aspects of quantum many-body physics, leading to confrontation of theory with experiment.
The specific systems probed include homogeneous and trapped Fermi gases, polarons, optical lattices, 
Fermi-Fermi and Bose-Fermi mixtures, lower-dimensional
systems, spin-orbit coupled gases, among several related settings~\cite{Dalfovo:1999,Bloch:2008,Giorgini:2008,Chin:2010,Enss:2011,Levinsen:2015,Chevy:2016}. 

In the study of quantum many-body physics an important feature of the system is the type of statistics that the particles obey. When the temperature is low enough such that the thermal de Broglie wavelength of the particles is of the same order as the interparticle spacing the particles are said to be indistinguishable. When this occurs in systems consisting of one species of particles, Fermi-Dirac statistics, in which quantum states can only be occupied by one particle and the wave function is antisymmetric under two-particle exchange, are needed for fermions and Bose-Einstein statistics, in which there is no limit to the number of particles occupying a state and the wave function is symmetric under two-particle 
exchange, for bosons. At larger temperatures, such systems follow Maxwell-Boltzmann statistics~\cite{Pathria}. 

A current frontier in cold-atom physics
is the study of many-component gases: the leading contenders have been $^{173}$Yb
and $^{87}$Sr ~\cite{Cazalilla:2014,DeSalvo:2010,Taie:2010,Fukuhara:2007,Tey:2010,Stellmer:2011,Pagano:2014,Zhang:2014,Heinze:2013,Cazalilla:2009,Boninsegni:2012,Cheng:2017}. 
The motivation behind such experiments is to use a large number of atoms $N$ distributed among different states. The
natural extension of this approach is to keep increasing the number of possible states among which the total number of 
particles is distributed. The extreme case of this scenario is when the number of states equals the number of particles, namely each component/species is placed
in a distinct state: the particles may still be strongly interacting with each other, but the fermionic (or bosonic) nature
of the underlying atom is no longer relevant. 
This is the problem we are interested in in this work, namely the study of \textit{quantum boltzmannons}, 
where quantum mechanics plays a significant role but quantum statistics doesn't. Since each particle is taken to be 
in a different quantum state, the particles are distinguishable, so they follow Maxwell-Boltzmann statistics even
at low temperature.

Cold atoms have proven to be a good laboratory for studying nucleonic matter~\cite{Gezerlis:2008,Carlson:2012,Gandolfi:2015,Horikoshi:2016} Thus, experiments with two species of cold Fermi gases probe the physics of strong pairing, which is
very similar to that of low-density neutron matter, found in the inner crusts of neutron stars. Similarly, the three-species quantum problem and the related area of Efimov physics have been of interest to both nuclear and atomic physicists~\cite{Naidon:2016,Braaten:2007}. The 
obvious extension, the four-species problem, is of direct relevance to all of nuclear physics, since nuclei on earth and
nucleonic matter in astrophysical settings are all made up of neutrons and protons (with two spin-projection states each). 
As more species are added to the problem, one can attempt to disentangle the effect of interactions from that of statistics.
This is analogous to the study of gauge theories using the $1/N$ expansion~\cite{Appelquist:1981}.
Importantly, one can expect that the very-many-species cold-atom problem may be experimentally probed in the 
not too distant future. 

Quantum Monte Carlo (QMC) is a term describing a family of powerful theoretical simulation techniques applied to several different
physical systems, including cold-atomic gases. QMC methods are typically non-perturbative and can probe both weak coupling
and strong coupling, 
at both zero temperature~\cite{Carlson:2003,Astrakharchik:2004,Stecher:2008,Forbes:2011,Bertaina:2011,Shi:2015,Galea:2016,Shi:2016} and finite temperature~\cite{Akkineni:2007,Burovski:2008,Bulgac:2008,Houcke:2012,Rubenstein:2012,Braun:2015,Anderson:2015,Yan:2016}. The particular QMC method used in this work is Path Integral Monte Carlo (PIMC). The Path integral Monte Carlo simulation techniques described and utilized in this work are suited to systems that obey Maxwell-Boltzmann statistics (boltzmannons) and can be used to calculate thermodynamic properties of systems composed of distinguishable particles
in the quantum regime. With this technique it is possible to calculate quantities such as the energy, pressure and specific heat (to name a few) of interacting particles at finite temperatures. The interactions focussed on in this work are those of hard spheres and hard cavities. These interactions have been a topic of interest in PIMC simulations, and other computational studies, in the past due to the fact that they can be handled with relative computational ease, and they provide a standard form for repulsive interactions between atoms~\cite{Kalos:1974}. 

An important aspect of studying statistical mechanical properties is the evaluation of the partition function. However, for a large system of interacting particles it is extremely difficult to evaluate the partition function directly. In PIMC, the partition function is evaluated using a path integral approach where thermal density matrices can be thought of as propagators over discretized imaginary time slices that form a path in coordinate space. Approximations are utilized to evaluate the thermal density matrices for the specific interactions found in the system being studied. For hard-sphere and hard-cavity interactions two approximations are commonly used, the Image Approximation (IA) \cite{Jacucci:1983} and another derived by Cao \& Berne (CB) \cite{CaoBerne:1992}. These approximations become more accurate as the path becomes more discretized and the number of imaginary time-slices increases, becoming exact in the limit of infinite time slices. However, increasing the number of time-slices causes an increase in computational time. Thus, a good measure for the effectiveness of an approximation is how quickly its simulation results converge as a function of number of time-slices. Crucially, boltzmannons do not suffer from the fermion-sign problem, so a non-perturbative PIMC calculation for this system is in principle exact (of course, one must still carefully study finite-size effects, time slice errors, and so on, as we do below). 

In this work, we perform an analytic calculation of the energy for a two-body system interacting via a hard-sphere, hard-cavity interaction and perform PIMC simulations for the same system using both the IA and the CB approximations at varying temperatures. We study the convergence of both approximations to the analytic value as a function of number of time-slices. Additionally, we carry out calculations of the energy, pressure, and specific heat of a system of $N$ hard-sphere particles under periodic boundary conditions using the CB density matrix. We investigate finite-size effects 
and at what value of $N$ do these finite-size effects drop away and the thermodynamic limit can be said to have been reached.    

\section{Path Integral Monte Carlo Method}
\label{sec:PIMC Method}
\subsection{Partition Function in the Path Integral Monte Carlo Formalism}

A fundamental quantity in a statistical mechanical description of a system is the thermal density matrix. The thermal density matrix is defined as:
\begin{equation}
\widehat{\rho} = e^{-\beta\widehat{H}} = \sum_i\ket{\psi_i}\bra{\psi_i}e^{-\beta E_i} 
\end{equation}

\noindent Where $\beta=1/k_BT$, $k_B$ is the Boltzmann constant, $T$ is the temperature, $\psi_i$ are the eigenstates of the system, and $E_i$ are the associated eigenenergies. 

One often wishes to compute the partition function because of its usefulness in deriving other thermodynamic quantities. The partition function is defined as the trace of the thermal density matrix. In the PIMC formalism the trace is performed in the position basis:

\begin{equation}
\label{eq: SimpleTrace}
Z = \textit{Tr}(\widehat{\rho}) = \int d\bold{R} \expval{e^{-\beta\widehat{H}}}{ \bold{R} } 
\end{equation}
where $\bold{R}$ represents the set of positions of all $N$ particles in the system, $\bold{R}={\bold{r}_1,\bold{r}_2,...,\bold{r}_N}$. The matrix element in the above integration cannot in general be calculated exactly for Hamiltonians of interacting many-body systems. To continue with the evaluation of the partition function one can expand the above integral 
using the two following relations:

\begin{equation}
\begin{split}
e^{-\beta\widehat{H}} &=e^{\frac{-\beta}{2}\widehat{H}}e^{\frac{-\beta}{2}\widehat{H}} \quad \\
 \quad \widehat{1}=&\int d\bold{R} \ket{\bold{R}}\bra{\bold{R}} 
\end{split}   
\end{equation}
By using these relations $M-1$ times, the partition function can be written as:
\begin{equation}
\label{eq:PathIntPart}
\begin{split}
Z = \int...\int d\bold{R}d\bold{R}_1d\bold{R}_2...d\bold{R}_{M-1}\bra{\bold{R}}e^{-\frac{\beta}{M} \widehat{H}}\ket{\bold{R}_1} \times \\ \times\bra{\bold{R}_1}e^{-\frac{\beta}{M} \widehat{H}}\ket{\bold{R}_2}...\bra{\bold{R}_{M-1}}e^{-\frac{\beta}{M} \widehat{H}}\ket{\bold{R}}
\end{split}
\end{equation}
It can now be seen where the analogy to the Feynman path integral can be made. The operator $e^{-\frac{\beta}{M} \widehat{H}}$ is analogous to the time-evolution operator that evolves the system between subsequent states $ \ket{R_i}$, except $it/\hbar$ is replaced with $\beta/M$. As a result, $\beta/M$ is the so-called `imaginary time' and Eq.~(\ref{eq:PathIntPart}) can be thought of as a path integral with $M$ imaginary time slices.    

To continue the derivation it is necessary to evaluate the intermediate density matrix elements that are being integrated over. To do this, the Trotter-Suzuki formula is used:
\begin{equation}
e^{(\widehat{A_1}+\widehat{A_2})/M} \approx e^{\widehat{A_1}/M}e^{\widehat{A_2}/M}
\end{equation}
where $M$ is taken to be large. Here, $\widehat{A_1}$ and $\widehat{A_2}$ are operators that do not necessarily commute. Applying this formula to the intermediate density matrices yields:
\begin{equation}
\bra{\bold{R}''}e^{-\tau\widehat{H}}\ket{\bold{R}'} \approx \bra{\bold{R}''}e^{-\tau\widehat{K}}e^{-\tau\widehat{V}}\ket{\bold{R}'} 
\end{equation}
where, again, $M$ is taken to be large. Also, $\widehat{H} = \widehat{K} + \widehat{V}$ where $\widehat{K}$ it the kinetic operator, $\widehat{V}$ is the potential operator and $\tau=\beta/M$. From Trotter-Suzuki we see $M$ must be taken to be very large in order for this expression to be near 
exact. It is now possible to evaluate the matrix element of both the exponentiated kinetic and potential operators separately:
\begin{equation}
\bra{\bold{R}''}e^{-\tau\widehat{K}}\ket{\bold{R}'}= \Big( \frac{Mm}{2\pi\hbar^2\beta} \Big)^{3N/2}\exp \Big[ \frac{-Mm}{2\hbar^2\beta} (\bold{R}''-\bold{R}')^2  \Big]
\end{equation}

\begin{equation}
\bra{\bold{R}''}e^{-\tau\widehat{V}}\ket{\bold{R}'}=\exp \Big[-\frac{\beta}{M}V(\bold{R}')\Big]~\delta(\bold{R}'' - \bold{R}')
\end{equation}
Now the path integral form of the partition function can be written as:
\begin{equation}
\begin{split}
Z = \int...\int d\bold{R}_1d\bold{R}_2...d\bold{R}_{M}\Big( \frac{Mm}{2\pi\hbar^2\beta} \Big)^{3NM/2}   \times \\ \times \exp \Big[ \frac{-Mm}{2\hbar^2\beta} \sum_{l=1}^M(\bold{R}_l-\bold{R}_{l+1})^2  \Big]\exp \Big[-\frac{\beta}{M} \sum_{l=1}^{M}V(\bold{R}_l)\Big]
\end{split}
\end{equation}
which is exact in the limit of $M$ going to infinity. It is important to realize that in the above expression the position state $\bold{R}_M$ is the original state of which the trace in Eq.~(\ref{eq: SimpleTrace}) is being performed over. Additionally, the path integral in the above expression begins and ends at the same state since the partition function is an integral over the diagonal thermal density matrix elements. As a consequence, $\bold{R}_{M+1} = \bold{R}_1$.    

It should also be noted that the manner with which the potential operator component of the density matrix is handled here is called the `primitive approximation'. The present work does not use this approximation in calculations, but it is convenient for introducing and deriving the PIMC technique \cite{Ceperley:1995,RungeChester:1988}.

\subsection{Calculating Thermodynamic Averages}

As mentioned above, the previous derivation of the partition function uses the primitive approximation to evaluate thermal density matrices. In practice, we evaluate density matrices as follows:
\begin{equation}
\bra{\bold{R}''}e^{-\tau\widehat{H}}\ket{\bold{R}'}=\bra{\bold{R}''}e^{-\tau\widehat{K}}\ket{\bold{R}'}\prod_{i,j}\widetilde{\rho}(\bold{r}''_{i,j},\bold{r}'_{i,j},\tau)
\end{equation}
where $\bold{r}''_{i,j}=\bold{r}''_{i}-\bold{r}''_{j}$ , $\bold{r}'_{i,j}=\bold{r}'_{i}-\bold{r}'_{j}$ and $\widetilde{\rho}(\bold{r}''_{i,j},\bold{r}'_{i,j},\tau)$ is the two-body density matrix, which has replaced the exponentiated potential operator. The two-body density matrix contains information about the interactions between the particles in the system. For the hard sphere and hard cavity interactions that are of interest in this work, there are well-known two-body density matrices (See Secs.~\ref{sec: 2Part} and \ref{sec: ManyPart}).

Now that the partition function of the system has been found (with the use of the appropriate two-body density matrix) thermodynamic observables can be calculated. In general, these observables are calculated as:
\begin{equation}
\label{eq: ThermOgen}
\left\langle \widehat{O} \right\rangle = \frac{1}{Z} \textit{Tr}(\widehat{O}\widehat{\rho}) 
\end{equation}
In the PIMC formalism Eq.~(\ref{eq: ThermOgen}) becomes:
\begin{equation}
\label{eq: GenExpect}
\begin{split}
\left\langle \widehat{O} \right\rangle =  \int d\bold{\mathcal{R}} O(\bold{\mathcal{R}})W(\bold{\mathcal{R}}) 
\end{split}
\end{equation}
where $\bold{\mathcal{R}} = \left\lbrace \bold{R}_1,\bold{R}_2,...,\bold{R}_M \right\rbrace$, which is referred to as the path, and $W(\bold{\mathcal{R}})$ can be thought of as a probability distribution of all possible paths written as: 
\begin{equation}
\begin{split}
W(\bold{\mathcal{R}}) &= \frac{1}{Z}\Big( \frac{Mm}{2\pi\hbar^2\beta} \Big)^{3NM/2}\exp \Big[ \frac{-Mm}{2\hbar^2\beta} \sum_{l=1}^M(\bold{R}_l-\bold{R}_{l+1})^2  \Big] \times \\ 
\times &\prod_{l=1}^{M}\prod_{i,j}\widetilde{\rho}(\bold{r}_{l,(i,j)},\bold{r}_{l+1,(i,j)},\tau)
\end{split}
\end{equation}
where $\bold{r}_{l,(i,j)}=\bold{r}_{l,i}-\bold{r}_{l,j}$ and $\bold{r}_{l+1,(i,j)}=\bold{r}_{l+1,i}-\bold{r}_{l+1,j}$.

For the calculation of specific observables the functional form of $O(\bold{\mathcal{R}})$ must be known. These functions are referred to as estimators and can be derived from the appropriate derivatives of the partition function. As an example, the energy of a system is given by:
\begin{equation}
\left\langle E \right\rangle = - \frac{\partial \ln Z}{\partial \beta}
\end{equation}
Carrying out this derivative gives the energy estimator:
\begin{equation}
\label{eq: EngEst}
\begin{split}
E(\bold{\mathcal{R}}) = \frac{3NM}{2\beta}-\frac{Mm}{2 \hbar^2 \beta^2 } \sum_{l=1}^M(\bold{R}_l-\bold{R}_{l+1})^2 - \\ 
- \sum_{l=1}^M \sum_{i,j}\frac{\partial \ln (\widetilde{\rho}(\bold{r}_{l,(i,j)},\bold{r}_{l+1,(i,j)},\tau))}{\partial \beta}
\end{split}
\end{equation}
wxhere $\Omega$ is the volume of the simulation box. An estimator for pressure can be derived in a similar manner. The average pressure of a system is given by:
\begin{equation}
\left\langle P \right\rangle = \frac{1}{\beta} \frac{\partial \ln Z}{\partial \Omega}
\end{equation}
leading to:
\begin{equation}
\label{eq: PrEst}
\begin{split}
P(\bold{\mathcal{R}}) = \frac{NM}{\beta \Omega}-\frac{Mm}{3 \hbar^2 \beta^2 \Omega } \sum_{l=1}^M(\bold{R}_l-\bold{R}_{l+1})^2 + \\ 
+ \frac{1}{\beta} \sum_{l=1}^M \sum_{i,j}\frac{\partial \ln (\widetilde{\rho}(\bold{r}_{l,(i,j)},\bold{r}_{l+1,(i,j)},\tau))}{\partial \Omega}
\end{split}
\end{equation}
With these estimators the average energy and pressure can be calculated by plugging Eqs.~(\ref{eq: EngEst}) and (\ref{eq: PrEst}) into the integral of Eq.~(\ref{eq: GenExpect}). However, these integrals cannot be evaluated analytically. Instead, a standard Metropolis algorithm is used to sample configurations from the set $\left\lbrace \bold{\mathcal{R}} \right\rbrace$ according to the probability distribution $W(\bold{\mathcal{R}})$. The estimators are then evaluated at each sampled configuration and the average is taken. Therefore the final expressions for the average energy and pressure of a system are:
\begin{equation}
\begin{split}
\left\langle E \right\rangle = \Big \langle \frac{3NM}{2\beta}-\frac{Mm}{2 \hbar^2 \beta^2 } \sum_{l=1}^M(\bold{R}_l-\bold{R}_{l+1})^2 - \\ 
- \sum_{l=1}^M \sum_{i,j}\frac{\partial \ln (\widetilde{\rho}(\bold{r}_{l,(i,j)},\bold{r}_{l+1,(i,j)},\tau))}{\partial \beta} \Big \rangle
\end{split}
\end{equation}

\begin{equation}
\begin{split}
\left\langle P \right\rangle = \Big \langle \frac{NM}{\beta \Omega}-\frac{Mm}{3 \hbar^2 \beta^2 \Omega } \sum_{l=1}^M(\bold{R}_l-\bold{R}_{l+1})^2 + \\ 
+ \frac{1}{\beta} \sum_{l=1}^M \sum_{i,j}\frac{\partial  \ln(\widetilde{\rho}(\bold{r}_{l,(i,j)},\bold{r}_{l+1,(i,j)},\tau))}{\partial \Omega} \Big \rangle
\end{split}
\end{equation}
where $\langle \cdots \rangle$ denotes an average over configurations sampled with the Metropolis algorithm. As mentioned previously, the number of time slices, or $M$, in the above expressions is an arbitrary parameter that can be set to any positive integer value. However, as discussed, the approximations that were required to derive these expressions require large $M$ to be accurate. As a result, calculations of the energy and pressure in PIMC simulations will converge, over increasing $M$, to the correct value \cite{Ceperley:1995,RungeChester:1988}. 

\section{Two-Particle Hard-Sphere \& Hard-Cavity System}
\label{sec: 2Part}
\subsection{Analytic Calculation of Energy}

An objective of this work was to test the effectiveness of two well-known approximations for two-body density matrices used in hard-sphere and hard-cavity interactions. To do this, we first analytically calculate the energy of a system consisting of two particles that have a hard-sphere radius of $\sigma$ and cannot be separated by a distance greater than a specified hard-cavity radius $r_{cav}$. Once this calculation was performed for various temperatures, PIMC simulations were also performed for the same system using both of the two-body density matrix approximations. The convergence of these simulations to the analytic results as a function of $M$ was then observed.       

The two-body Schr\"odinger equation when the potential is a function of the distance between the particles can be separated into the following differential equations:
\begin{equation}
\label{eq:CoM_SE}
-\frac{\hbar^2}{2M}\nabla^2_R\psi=E_M\psi
\end{equation}

\begin{equation}
\label{eq:ReducedM_SE}
-\frac{\hbar^2}{2\mu}\nabla^2_r\psi+V(|\bold{r}_1-\bold{r}_2|)\psi=E_\mu \psi 
\end{equation}
where we use $|\bold{r}_1-\bold{r}_2|$ and $r$ interchangeably.
Our task has now been separated into two problems: one is a free particle of mass $M = 2m$ in the centre of mass position Eq.~(\ref{eq:CoM_SE}), and the other is a particle of reduced mass $\mu = m/2$ whose radial component is that of the separation distance in the original problem Eq.~(\ref{eq:ReducedM_SE}). $E_M$ denotes the centre of mass energy, and $E_\mu$ is the separation distance energy. 

To solve for the expectation value of the energy at a finite temperature, the energy levels of the system must be solved for and then averaged using Boltzmann statistics. 

The centre of mass energy is continuous since it is a free particle and can be calculated as: 
\begin{equation}
\begin{split}
\langle E_M \rangle &=\frac{\Omega m^{3/2}}{\sqrt{2}\pi^2 \hbar^3 Z} \int_{0}^{\infty}E^{3/2}e^{-\beta E}dE \\
Z &=\frac{\Omega m^{3/2}}{\sqrt{2}\pi^2 \hbar^3}\int_{0}^{\infty}E^{1/2}e^{-\beta E}dE
\end{split}
\end{equation}
which can now be solved for a general inverse temperature $\beta$. 

The potential for the two-particle system we are studying, $V(|\bold{r}_1-\bold{r}_2|)$, is defined in the following way:
\[   
V(|\bold{r}_1-\bold{r}_2|) = 
     \begin{cases}
       \text{0} &\quad\text{if} \quad \sigma \le |\bold{r}_1-\bold{r}_2| \le r_{cav}\\
       \infty &\quad\text{otherwise} \\
     \end{cases}
\]
This leads to the following differential equation for the radial component of the wavefuntion in the separation distance:
\begin{equation}
\begin{split}
\frac{d^2R}{dr^2}+\frac{2}{r}\frac{dR}{dr} &+ ( k^2-\frac{l(l+1)}{r^2}) R = 0 \\ 
k &= \frac{\sqrt{2mE_\mu}}{\hbar}
\end{split}
\end{equation}
The solutions to this differential equation are the spherical Bessel functions of the first and second kind, therefore the radial wavefunctions are taken to be:
\begin{equation}
R_{l}(r) = Aj_l(kr)+Bn_l(kr) 
\end{equation}
where the $j_l$'s are the first kind and the $n_l$'s are the second. The $k$ values are solved for by imposing the boundary conditions of the problem: $R(\sigma)=0$ and $R(r_{cav})=0$. This results in the following transcendental equation that $k$ must satisfy:
\begin{equation}
j_l(r_{cav}k)-\frac{j_l(\sigma k)}{n_l(\sigma k)}n_l(r_{cav}k)=0
\end{equation}
Once the solutions for $k$ have been determined, the energy levels for the reduced mass component of the energy are given by:
\begin{equation}
E_{\mu,l,i}=\frac{\hbar^2k_{l,i}^2}{2\mu}
\end{equation}
where $l$ is the $l$'th spherical Bessel function and $i$ is $i$'th $k$ value associated with the $l$'th spherical Bessel function. According to Boltzmann statistics, the expectation value of the energy becomes:
\begin{equation}
\begin{split}
\langle E_\mu \rangle =\frac{1}{Z}\sum_{l=0}^{\infty}&\sum_{m_l=-l}^{l}\sum_{i=1}^{\infty}E_{\mu,l,i}e^{-\beta E_{\mu,l,i}} \\
Z = \sum_{l=0}^{\infty}&\sum_{m_l=-l}^{l}\sum_{i=1}^{\infty}e^{-\beta E_{\mu,l,i}}
\end{split}
\end{equation}
where $m_l$ is the regular magnetic quantum number introduced in the 3D Schr\"odinger equation solved in spherical coordinates. Since the potential has no angular dependence $m_l$ introduces a $2l+1$ degeneracy in the energy levels.   

Now the total energy can be calculated as:
\begin{equation}
\langle E \rangle = \langle E_M \rangle + \langle E_\mu \rangle 
\end{equation}

\subsection{PIMC Calculation of Energy}

\begin{figure}[b!]
\centering
\includegraphics[width=0.45\textwidth]{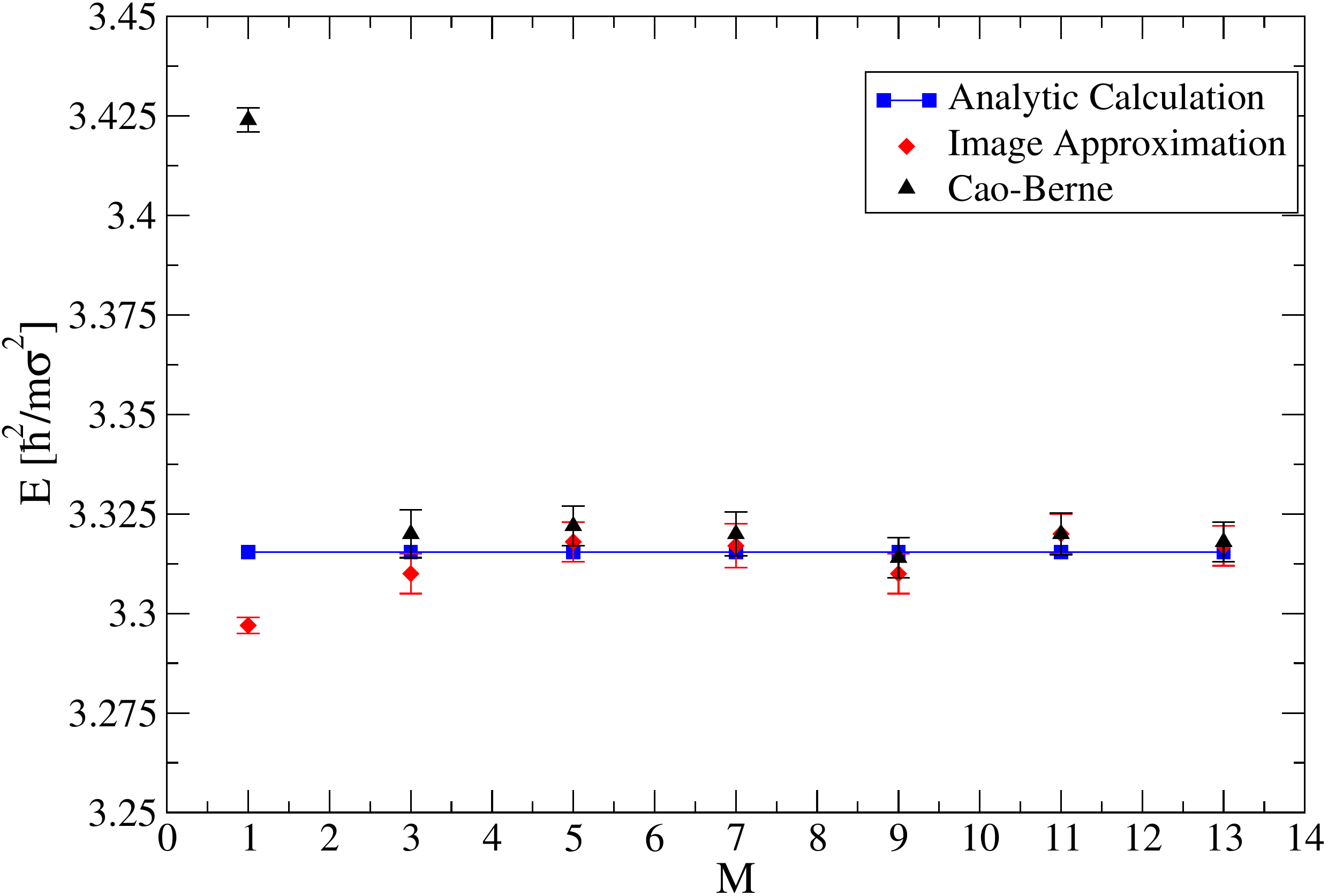} 
\caption{PIMC results for the two particle hard-sphere, hard-cavity system using both the IA and CB density matrix approximations compared to analytic results. The temperature of the system is  $T (\hbar^2/m\sigma^2k_B)^{-1} =1.0$ and $r_{cav}/\sigma=6$. Convergence of the PIMC results to the analytic energy occurred at approximately the same rate for both density matrices.}
\label{fig:2PartHSHC1T}
\end{figure}

The two-body density matrices used in this work are the IA and the CB density matrix. The Image Approximation 
is a simple way of meeting
the requirement of going to zero as $r$ goes to $\sigma$ or $r_{cav}$. On the other hand, the CB two-body density matrix is based on the partial-wave scattering solution of hard-sphere potentials and is a more general formula (which reduces
to the IA one under specific conditions). Because of this, it is expected the CB density matrix will yield better convergence. The functional forms of the two-body density matrices used in this work are given as: \cite{CaoBerne:1992}

\begin{figure}[!b]
\centering
\includegraphics[width=0.45\textwidth]{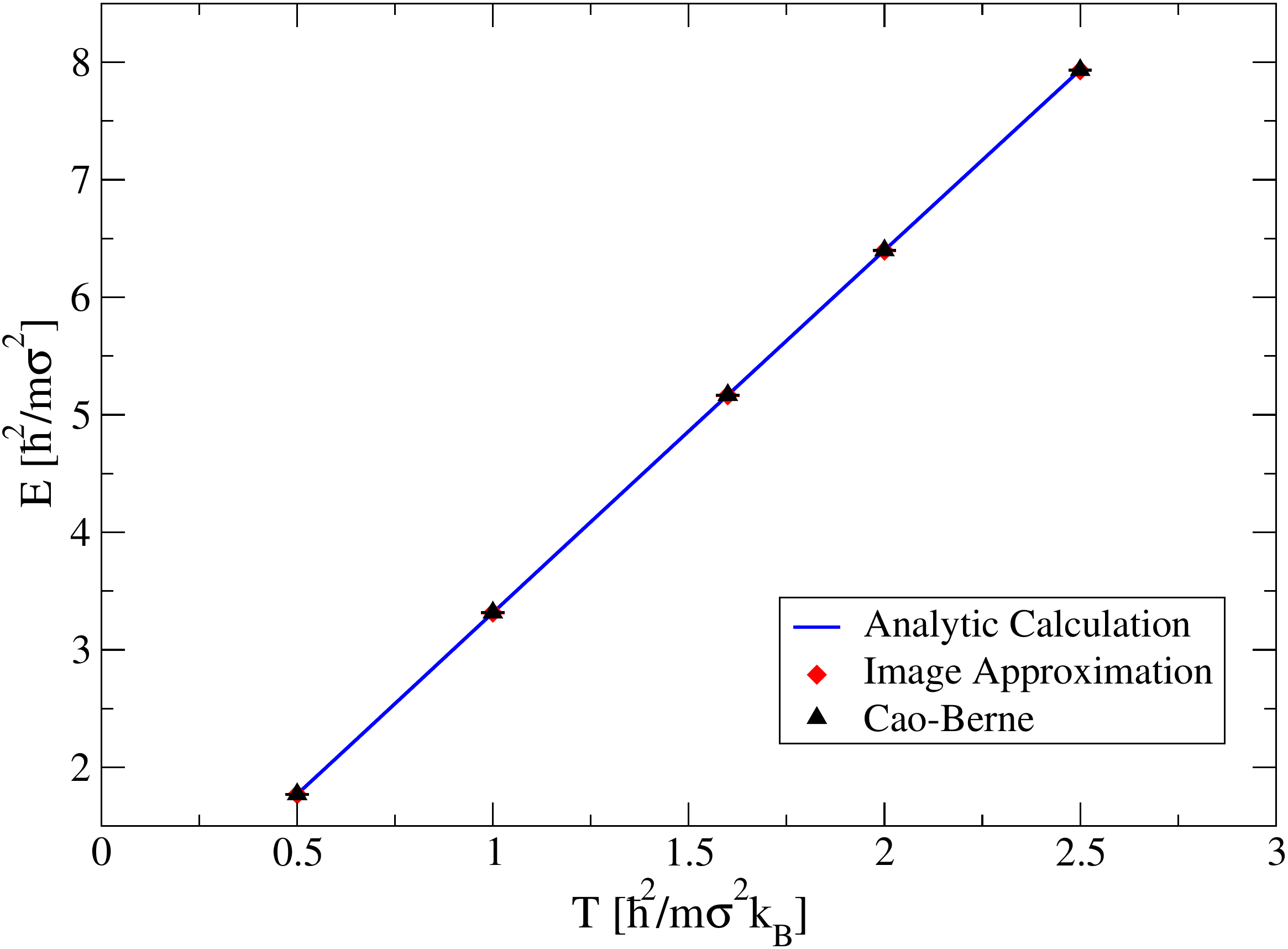} 
\caption{PIMC results of the energy for two particle hard-sphere, hard-cavity interaction for both Image Approximation and CB thermal density matrices.}
\label{fig:2PartHSHCEvsT}
\end{figure}

\begin{equation}
\label{Eq:IAtwobody}
\begin{split}
\widetilde{\rho}_{IA}(\textbf{r},\textbf{r}')=(1-\exp[- (M m/ \beta \hbar^2)(\textit{r}-\sigma)(\textit{r}'-\sigma)]) \times \\ 
\times (1-\exp(- (M m/ \beta \hbar^2)(r_{cav}-\textit{r})(r_{cav}-\textit{r}')])
\end{split} 
\end{equation}

\begin{equation}
\label{Eq:CBtwobody}
\begin{split}
\widetilde{\rho}_{CB}(\textbf{r},\textbf{r}')&=(1-\frac{\sigma(\textit{r} + \textit{r}') - \sigma^2 }{\textit{r} \textit{r}'} \times \\
\times &\exp[-(M m/ 2\beta \hbar^2)(\textit{r}-\sigma)(\textit{r}'-\sigma)(1+cos\chi)]) \times \\
\times &(1-\frac{2\textit{r}_{cav}-\textit{r}}{\textit{r}} \times \\
\times &\exp[- (M m/ 4 \beta \hbar^2)((\textbf{r}_{-1} + \textbf{r}')^2 - ( \textbf{r} - \textbf{r}')^2 ]) \\ 
\end{split}
\end{equation}

\begin{equation}
\textbf{r}_{-1} = (\textit{r}-2r_{cav}) \boldsymbol{\hat{r}}
\end{equation}
where we have defined $\textit{r}$ to always be the larger of the two vector magnitudes, i.e. $\textit{r} \geq \textit{r}'$ and $\chi$ is the angle 
between \textbf{r} and \textbf{r}$'$. Both  $\widetilde{\rho}_{IA}$ and $\widetilde{\rho}_{CB}$ are set to zero if $\textit{r}$ or $\textit{r}'$ are less than $\sigma$ or greater than $r_{cav}$.

Calculations of the energy of the system were performed at five different temperatures for increasing number of time slices. The results are presented in reduced units where $\sigma$ is the unit of length and $\hbar^2/m\sigma^2$ is the unit of energy.   

Convergence studies for the two density matrices using the two-particle calculation were carried out at varying temperatures, $T(\hbar^2/m\sigma^2k_B)^{-1}=0.5,1.0,1.6,2.0,2.5$. The results for the convergence study at $T(\hbar^2/m\sigma^2k_B)^{-1}=1.0$ are presented in Fig.~\ref{fig:2PartHSHC1T}. It was found that in all cases of varying temperature the general behaviour of the PIMC results vs temperature as a function of time slices remains fairly constant. For this system, PIMC simulations converge rather quickly to the analytic result regardless of the density matrix approximation that is used. It could be argued the Image Approximation density matrix gives slightly quicker convergence over the CB density matrix, but the difference is fairly insignificant. 

PIMC results for the energy vs temperature for both thermal density matrices and analytic results are plotted in Fig.~\ref{fig:2PartHSHCEvsT}. Both thermal matrices are in close agreement with each other and both agree very well with the linear relation seen from the analytic results over all temperatures studied.

Even though a significant difference between the convergence of the density matrices was not observed, it was decided that the CB density matrix should be used in the many body system calculations moving forward. This is because, as mentioned earlier, the CB form was more rigorously derived based on the partial-wave scattering solution for the hard-sphere interaction.

\section{Many Hard-Sphere Particle System }
\label{sec: ManyPart}

\subsection{Non-Interacting boltzmannon Gas}

\begin{figure}[t!]
\centering
\includegraphics[width=0.45\textwidth]{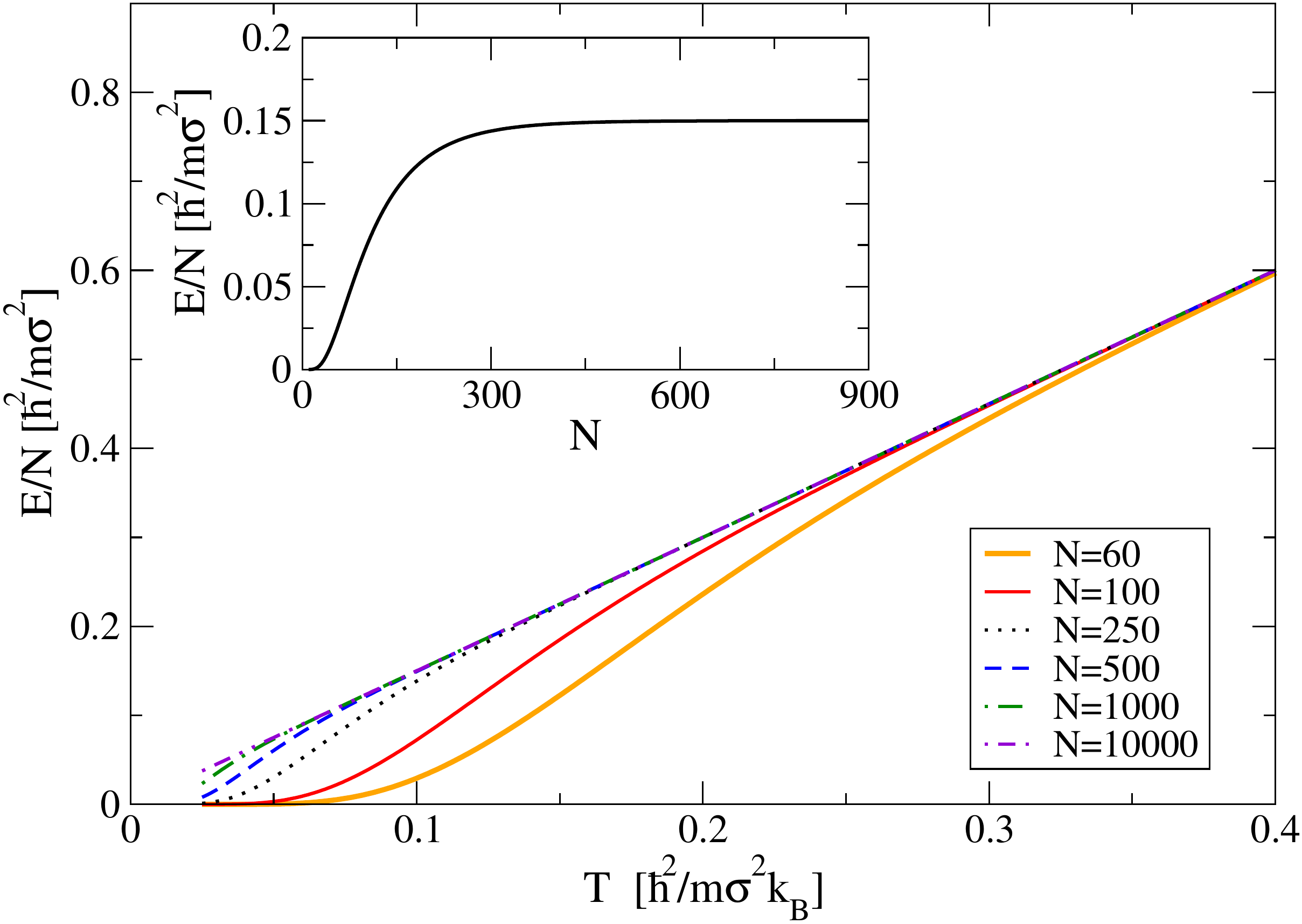} 
\caption{Energy per particle of non-interacting gas of distinguishable particles vs temperature for increasing particle number. At small values of $N$ finite-size effects cause a deviation from the expected $E/N=\frac{3}{2}k_BT$; as $N$ is increased, the expected behaviour is observed at lower $T$. Inset: Energy per particle of non-interacting gas vs $N$ at $T(\hbar^2/m\sigma^2k_B)^{-1}=0.1$, $n \sigma^3=0.2063$. As the particle number rises, $E/N$ approaches a constant value.}
\label{fig:FreeGasEnergy_2063}
\end{figure}  

In this section we discuss the calculation of the energy and specific heat of a system composed of many non-interacting boltzmannons under periodic boundary conditions. Periodic boundary conditions (PBC) in simulations are meant to approximate an infinite system in which a relatively small simulation box is repeated in all directions, and particles that leave one side of the box re-enter from the opposite side. PBC can be mathematically expressed as:
\begin{equation} 
\psi(x,y,z) = \psi(x\pm L_x,y\pm L_y,z\pm L_z)
\end{equation}
where $L_{x,y,z}$ are the lengths of the simulation box in the $x,y,z$ directions. Often times, and in the case of our simulations, the geometry of the simulation box is taken to be a cube, therefore $L_x=L_y=L_z=L$.

\begin{figure}[t!]
\centering
\includegraphics[width=0.45\textwidth]{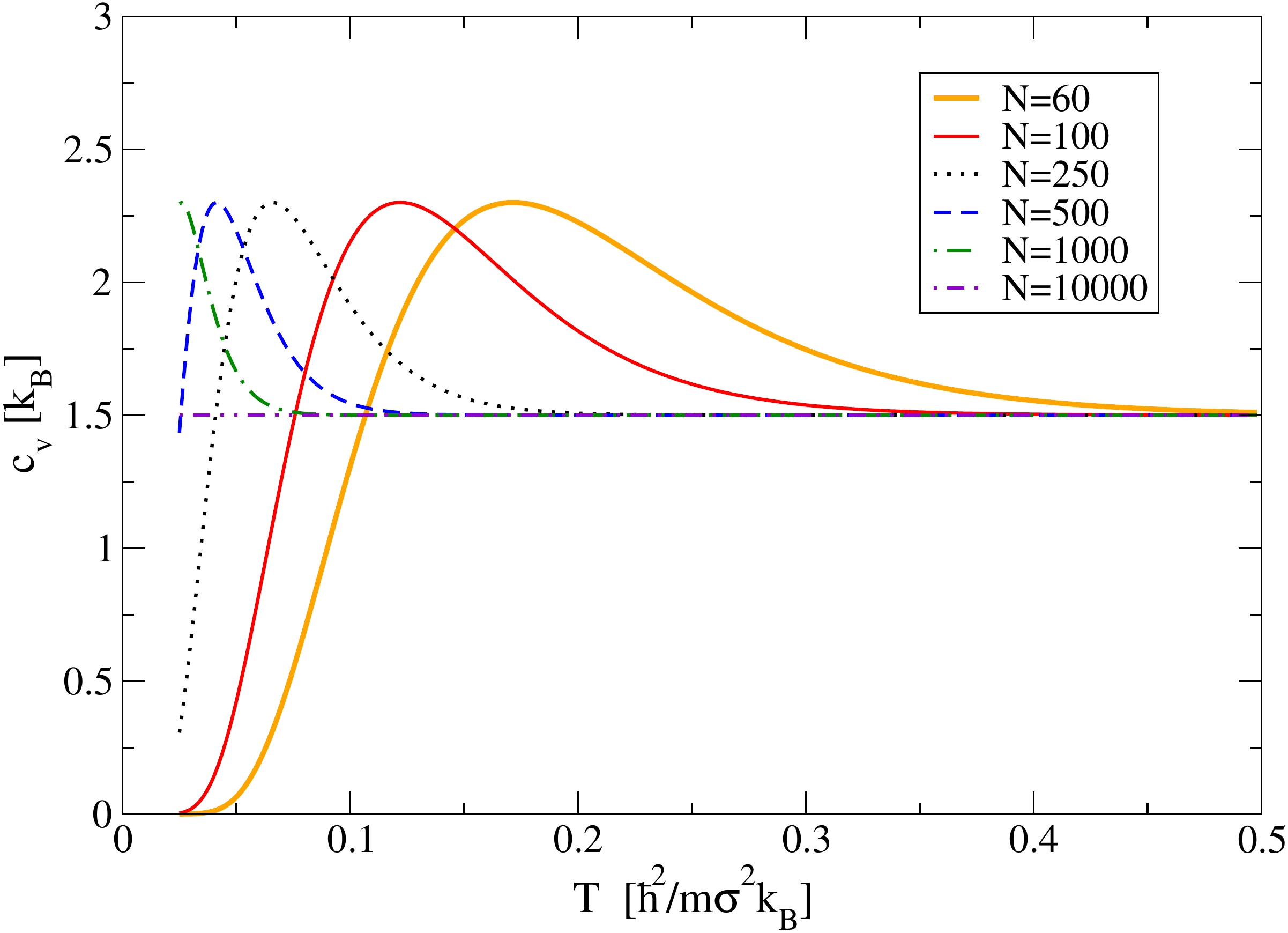} 
\caption{Specific heat of non-interacting gas of distinguishable particles vs temperature for increasing particle number, $n \sigma^3=0.2063$. At small values of $N$ a spike observed in $c_v$ at lower temperatures; as $N$ is increased, this spike becomes smaller and is seen at lower $T$ as finite-size effects fall away.}
\label{fig:FreeGasSpecificHeat_2063rho}
\end{figure}

PBC are useful as a simulation can be performed with a computationally manageable number of particles (10-1000), but can simulate properties of macroscopic systems at the thermodynamic limit when $N\rightarrow\infty$ and $\Omega \rightarrow\infty$, while the number density $n=N/\Omega$ is a constant. An issue with PBC is inaccuracies which are a result of finite-size effects. Finite-size effects are a result of the system within the simulation box being too small (i.e. too few particles). Finite-size effects can be observed by increasing the particle number and volume while maintaining a constant density and observing if intensive properties (i.e. independent of particle number) change. An example of finite-size effects can be seen in calculations of energy per particle and specific heat of a system of non-interacting boltzmannons. To begin, the eigenvalues of energy for a system of non-interacting particles under PBC are given as:
\begin{equation}
\begin{split}
E_n &= \frac{\hbar^2}{2m} |\textbf{k}_\textbf{n}|^2 \\
\textbf{k}_\textbf{n} = \frac{2 \pi}{L}&(n_x \boldsymbol{\hat{x}}+n_y \boldsymbol{\hat{y}}+n_z \boldsymbol{\hat{z}})
\end{split}
\end{equation}
where $L$ is the length of our simulation box and $n_x,n_y,n_z$ are integers. The energy of the system at a specific temperature is calculated by averaging the eigenenergies over the usual Maxwell-Boltzmann factor $e^{-\beta E_n}$. This calculation was performed at a constant density of $n \sigma^3=0.2063$ at increasing particle number values. The results are plotted in Fig.~\ref{fig:FreeGasEnergy_2063}. The expected behaviour of a non-interacting gas that obeys Boltzmann statistics is given by the equipartition theorem. For a 3D system where particles only have translational degrees of freedom, energy is related to temperature by:   
\begin{equation} 
\frac{E}{N} = \frac{3}{2}k_BT  
\end{equation}

In Fig.~\ref{fig:FreeGasEnergy_2063} it can be seen that as the particle number is decreased, finite-size effects cause a deviation from the equipartition theorem at increasingly higher temperatures. $E/N$ is plotted against $T$ for $N = (60, 100,250,500,1000,10000)$ and it can be seen that as $N$ increases, finite-size effects fall away and the results eventually converge. Finite-size effects are more prominent at lower $T$. For $N=100$ the results only begin to match what is expected at $T \approx 0.3$, which is drastically improved by the increase of the system size to $N=250$, and as $N$ is further increased, the expected result is found at lower and lower temperatures. The inset in Fig.~\ref{fig:FreeGasEnergy_2063} shows the reduction of finite-size effects at $T(\hbar^2/m\sigma^2k_B)^{-1}=0.1$ as $N$ is increased and $E/N$ eventually reaches the expected 0.15~$(\hbar^2/m\sigma^2)$, and then stays constant as $N$ is further increased. 

The specific heat of this system was calculated by the derivative of $E/N$ with respect to T at all values of $N$ previously used. Fig.~\ref{fig:FreeGasSpecificHeat_2063rho}  
again shows the convergence to the equipartition theorem at lower and lower $T$ as $N$ is increased. At lower $T$, $c_v$ increases from the expected values of 1.5~$k_B$, as predicted by the equipartition theorem, and then quickly decreases as $T$ is further decreased. This behaviour becomes less and less prominent as $N$ is increased, and $c_v$ eventually converges to the expected value.

Because of finite-size effects, care must be taken to ensure simulations are being performed with an adequate system size such that the results are representative of the thermodynamic limit. In the next section we explore at what system size the thermodynamic limit is reached for simulations of hard-sphere boltzmannons using the CB thermal density matrix.

\subsection{Hard-Sphere boltzmannons}

\begin{figure}[t!]
\centering
\includegraphics[width=0.45\textwidth]{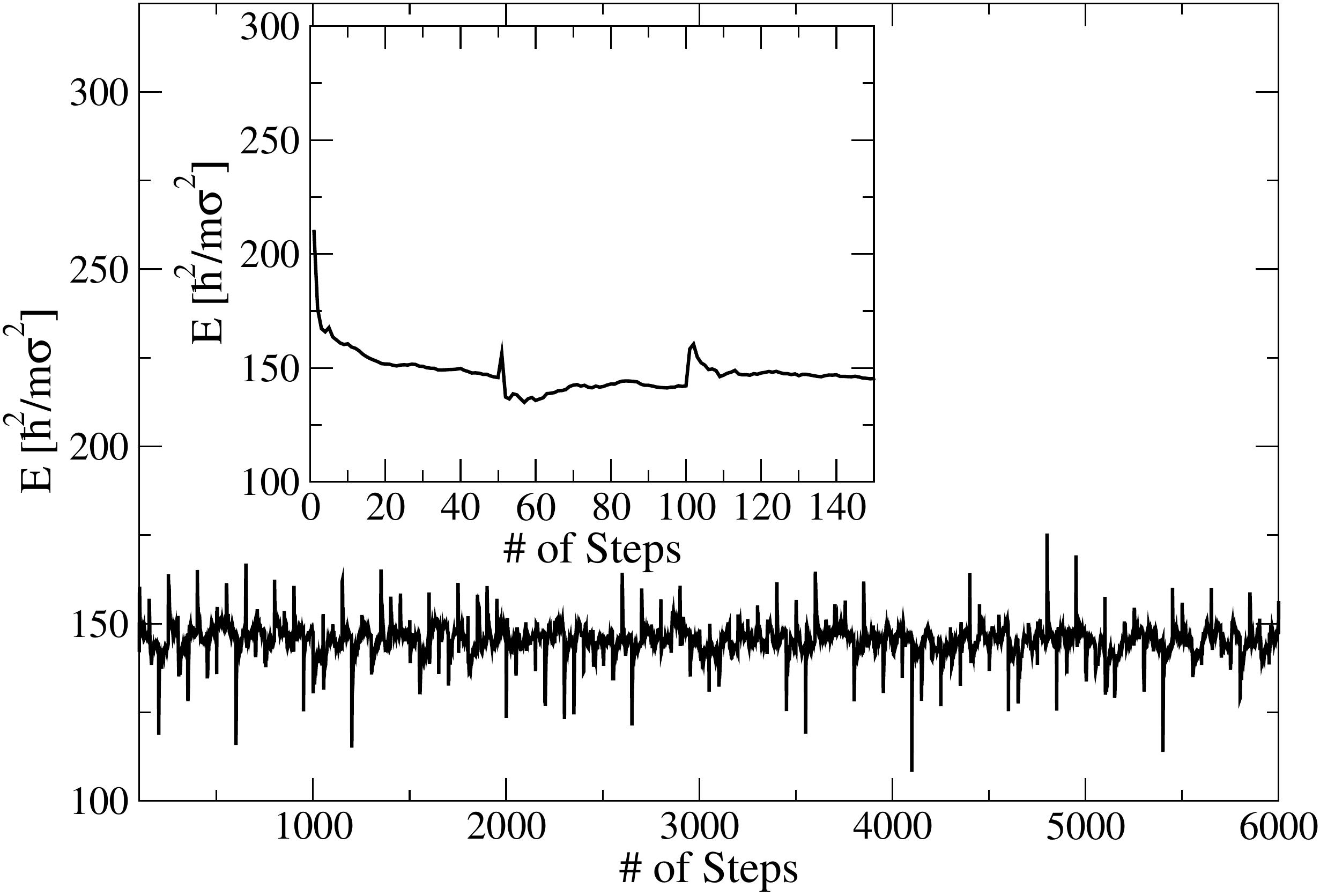} 
\caption{Energy of 20-particle system produced by PIMC vs the number of steps in the Metropolis algorithm. When taking the average of these results equilibration time must be taken into account. This calculation was performed at $T(\hbar^2/m\sigma^2k_B)^{-1}=1.0$, $n \sigma^3=0.2063$ and $M=41$. Inset: A closer look at the equilibration of the system over the early steps.}
\label{MonteCarloSteps20Part_2063rho_1T_41ntau}
\end{figure}

In this section we study systems of hard sphere particles using the PIMC methods described in the previous sections. As opposed to the 
two-particle calculation, this system does not have hard-cavity interactions present, as a result the CB two-body density matrix takes the 
form:
\begin{equation}
\label{Eq:CBtwobodyNoCav}
\begin{split}
\rho_{CB}(\textbf{r},\textbf{r}')&=(1-\frac{\sigma(\textit{r} + \textit{r}') - \sigma^2 }{\textit{r} \textit{r}'} \times \\
\times &\exp[-(M m/ 2\beta \hbar^2)(\textit{r}-\sigma)(\textit{r}'-\sigma)(1+cos\chi)])
\end{split}
\end{equation}
Fig. \ref{MonteCarloSteps20Part_2063rho_1T_41ntau} shows the energy for a simulation of a 20-particle system with 
$T(\hbar^2/m\sigma^2k_B)^{-1}=1.0$, $n \sigma^3=0.2063$ and $M=41$ at each configuration sampled by the Metropolis algorithm. An important 
detail of these simulations is to account for the equilibration time, which can be seen in the inset of Fig.~\ref{MonteCarloSteps20Part_2063rho_1T_41ntau}. When evaluating averages, 
one should only include values taken after the system has equilibrated.

We perform our calculations for the energy per particle and pressure using the PIMC method at varying temperatures and particle numbers while maintaining a constant density. As stated earlier, intensive properties such as the energy per particle or pressure are not affected by the number of particles in the system, but rather the number density ($N/\Omega$), that is, a system of varying particle number, but constant density, will have a constant $E/N$ and $P$. In our simulations it is expected that as the number of particles in our simulation box increases, $E/N$ and $P$ will vary at small values of $N$, but will eventually reach a constant value, analogously to the non-interacting case.

\begin{figure}[!t]
\centering
\includegraphics[width=0.45\textwidth]{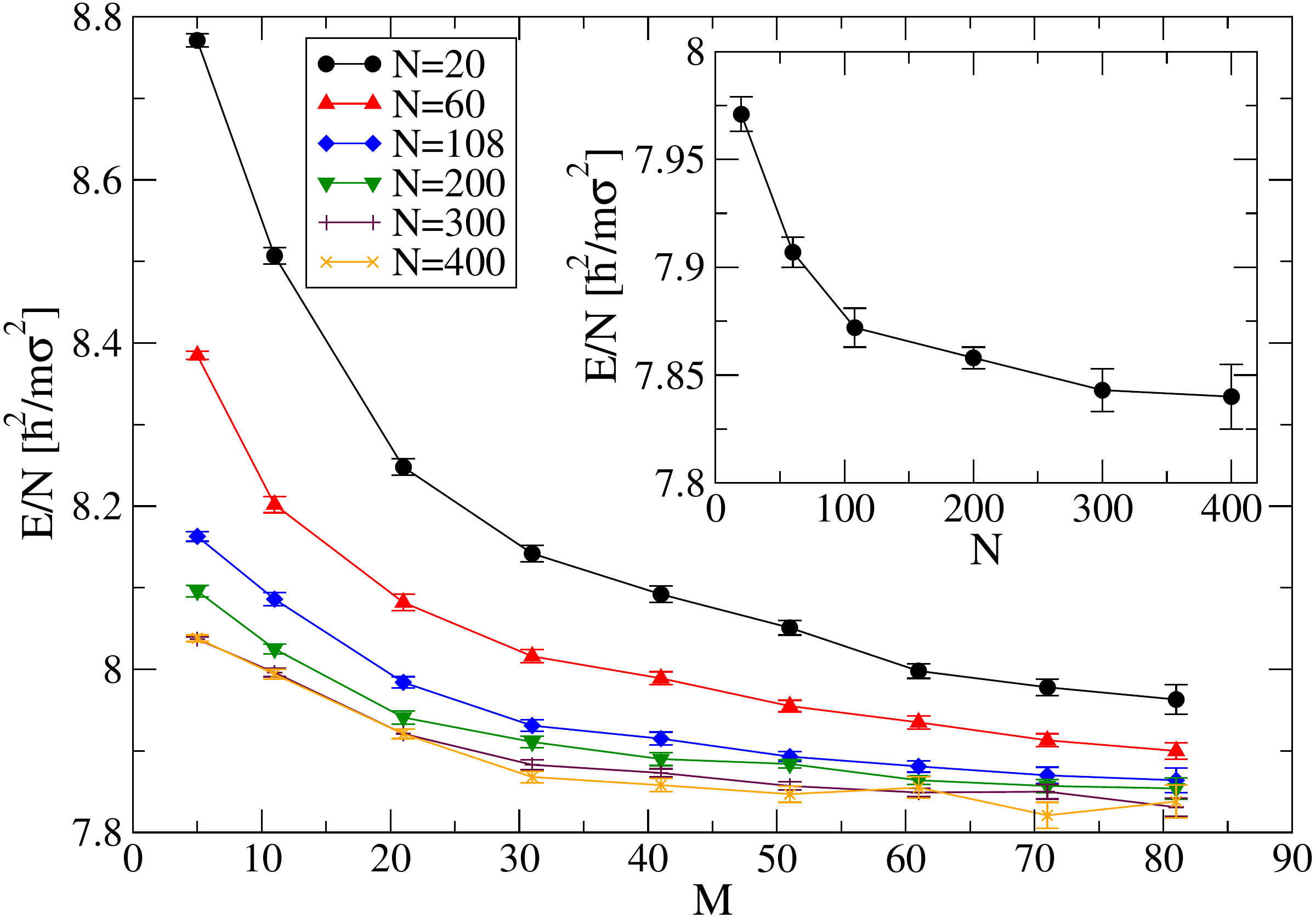}  
\caption{PIMC results of energy for a system of hard-sphere particles under periodic boundary conditions at varying particle number for $T(\hbar^2/m\sigma^2k_B)^{-1}=2.0$ and $n \sigma^3=0.2063$. For each value of $N$ the value of $M$ was increased until $E/N$ sufficiently converged. $N$ was increased until the thermodynamic limit was reached; this was found to be in the $N=300-400$ range. Inset: The converged values of $E/N$ for all values of $N$ at $T(\hbar^2/m\sigma^2k_B)^{-1}=2.0$.}
\label{fig:CaoBerne2TEnergyPIMC}
\end{figure} 

\begin{figure}[!b]
\centering
\includegraphics[width=0.45\textwidth]{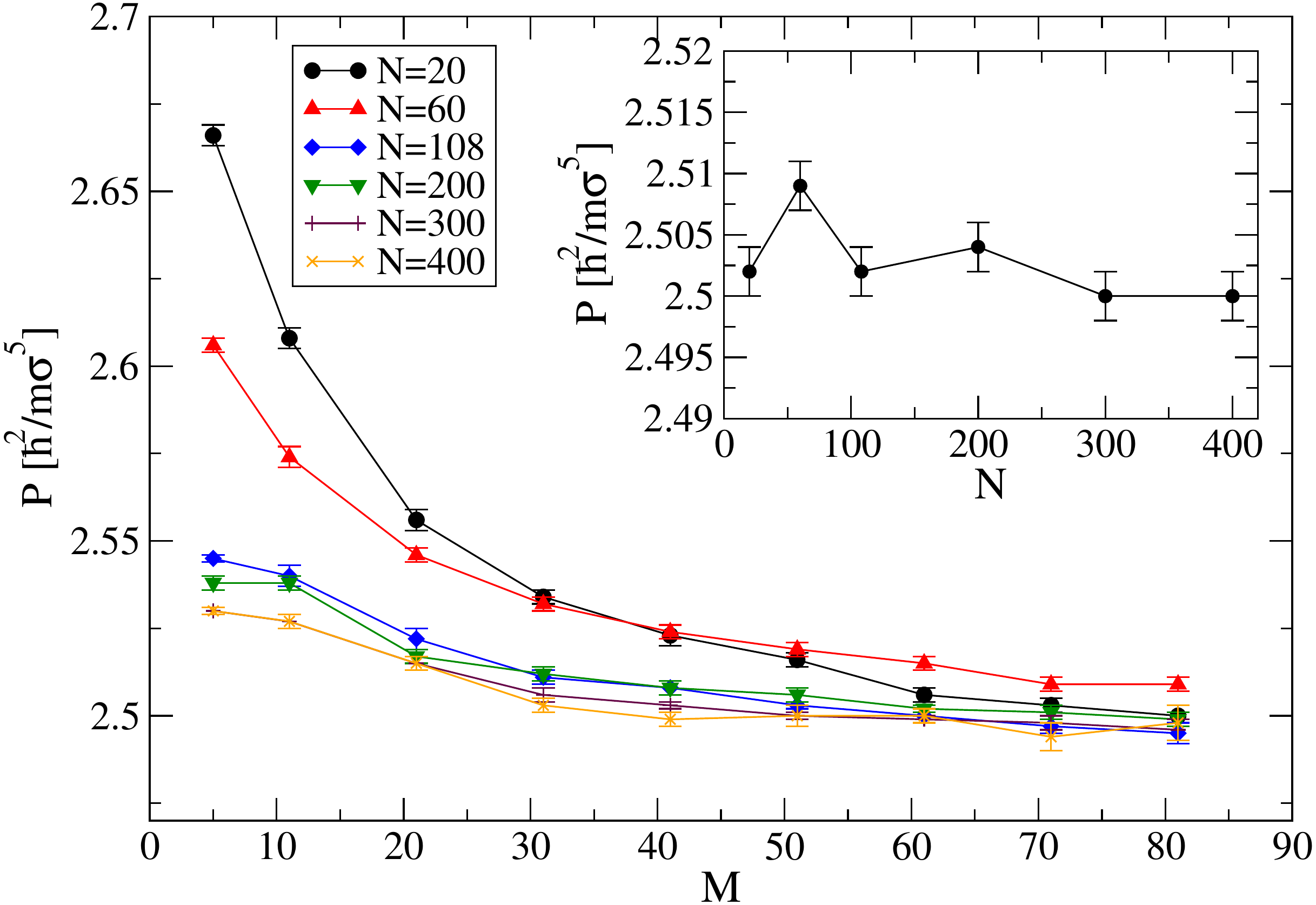}  
\caption{PIMC results of pressure for a system of hard-sphere particles under periodic boundary conditions at varying particle number for $T(\hbar^2/m\sigma^2k_B)^{-1}=2.0$ and $n \sigma^3=0.2063$. For each value of $N$ the value of $M$ was increased until $P$ sufficiently converged. $N$ was increased until the thermodynamic limit was reached, this was found to be in the $N=300-400$ range. Inset: The converged values of $P$ for all values of $N$ at $T(\hbar^2/m\sigma^2k_B)^{-1}=2.0$}
\label{fig:CaoBerne2TPressurePIMC}
\end{figure}   

In our PIMC calculations, particle number was increased until the results for energy per particle and pressure reached a final value. These calculations were performed at varying temperatures ($T(\hbar^2/m\sigma^2k_B)^{-1}=0.5,1.0,1.5,2.0,2.5,3.0$) and varying particle numbers ($N=20,60,108,200,300,400$) at a density $n \sigma^3=0.2063$. This density was chosen such that the Wigner-Seitz radius $r_0 = (3/4 \pi n)^{1/3} \approx 1.05 \sigma$, which ensures the system is strongly interacting via the hard-sphere potential. At all the above listed points in the ($N,T$) plane, simulations were performed where $M$ was increased in steady increments until the results no longer varied by a statistically significant amount. When the converged values of $E/N$ and $P$ at specific values of $N$ no longer vary with increasing $N$, the thermodynamic limit has been reached.

\begin{figure}[!t]
\centering
\includegraphics[width=0.45\textwidth]{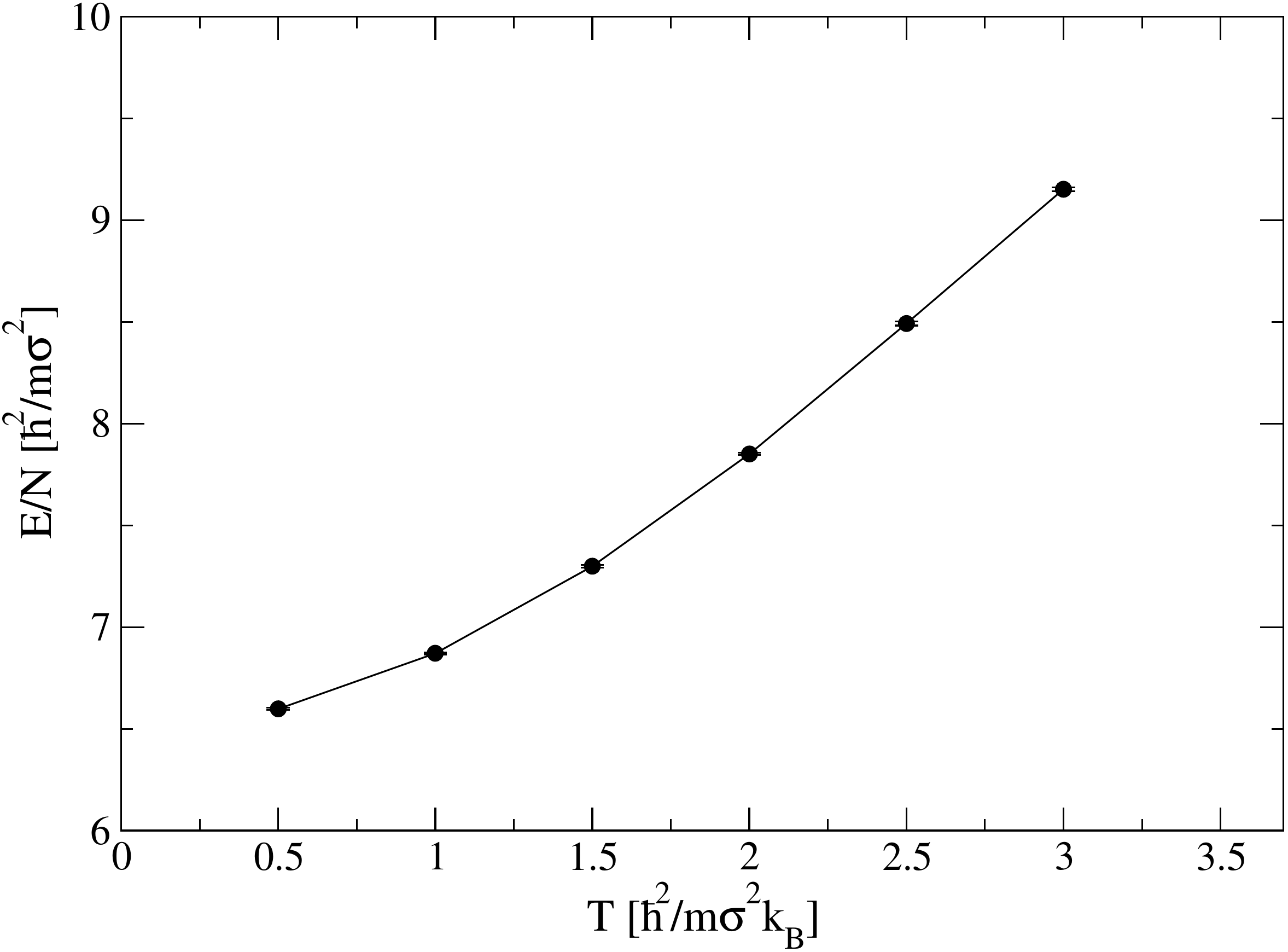} 
\caption{Energy per particle of system of hard-sphere particles at the thermodynamic limit for density $n \sigma^3=0.2063$. At higher temperatures $E/N$ is linear in $T$, while this relationship falls off as temperature is decreased. This shows that at high $T$ this system reaches the classical limit, while at low $T$ quantum effects take over.}
\label{fig:CaoBerneEperNVsT}
\end{figure}

Fig.~\ref{fig:CaoBerne2TEnergyPIMC} shows our calculation of $E/N$  for $T(\hbar^2/m\sigma^2k_B)^{-1}=2.0$. The general trends observed at $T(\hbar^2/m\sigma^2k_B)^{-1}=2.0$ are seen at all temperature values we studied. As the particle number is increased, $E/N$ decreases monotonically at all values of time slices, eventually settling at a constant value. It was also observed that convergence in $M$ was slower at lower values of $N$. This can be be seen in Fig.~\ref{fig:CaoBerne2TEnergyPIMC} where the value $M$ needed to be taken to 81 in order to observe convergence in $N=20,60$ and only $M=61$ was needed for the larger values of $N$. The inset of Fig.~\ref{fig:CaoBerne2TEnergyPIMC} shows the final converged values of $E/N$ at varying $N$ for $T(\hbar^2/m\sigma^2k_B)^{-1}=2.0$, where the monotonic decreasing of the energy to a constant value can be clearly observed. 

\begin{figure}[b]
\centering
\includegraphics[width=0.45\textwidth]{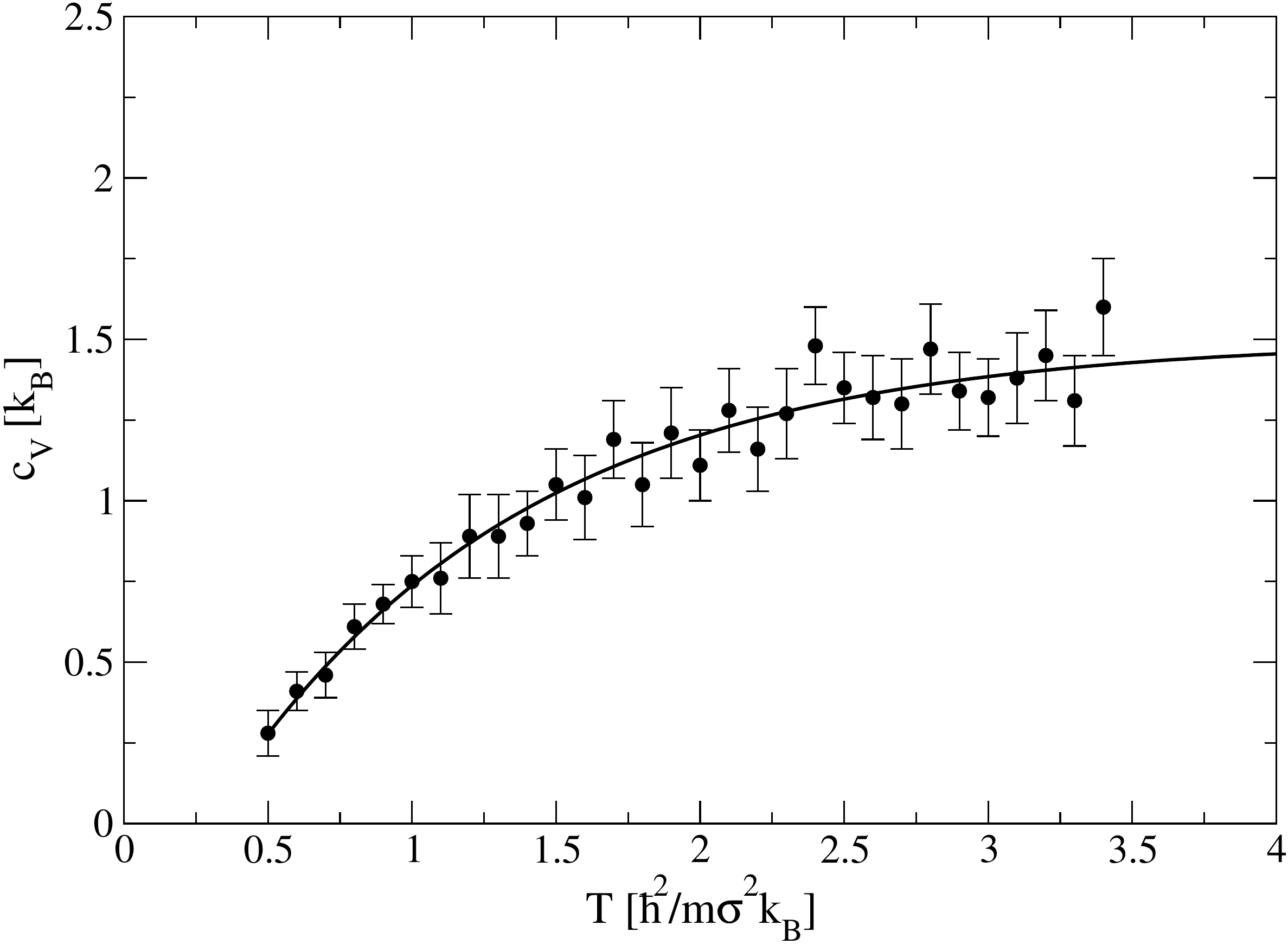} 
\caption{Specific heat of system of hard-sphere particles at the thermodynamic limit with density $n \sigma^3=0.2063$. As $T$ is raised $c_v$ approaches the 1.5~$k_B$ limit predicted by the equipartition theorem. In the range plotted $c_v$ decreases with $T$, as a consequence of the third law of thermodynamics.}
\label{fig:CaoBerne300PartCv}
\end{figure}   

The pressure of the same system is plotted in Fig.~\ref{fig:CaoBerne2TPressurePIMC}. The pressure also follows the general trend of decreasing  as the number of particles is increased, eventually converging to a final value, however unlike $E/N$, this does not occur monotonically at all temperatures, as can be seen in the higher time slice values of $N=20$ and $60$ for the $T(\hbar^2/m\sigma^2k_B)^{-1}=2.0$ case where the pressure increases between the two values of $N$. It can also be observed that convergence in $M$ is slower at smaller particle numbers, similar to the $E/N$ calculations. The converged values of $P$ for each value of $N$ are shown in the inset of this plot. Again, unlike the $E/N$ case we do not see a monotonically decreasing convergence to a final value, but instead an oscillatory convergence. After the above analysis was performed at each of the previously listed temperatures, it was found that a system size of $N=300-400$ was sufficient to have reached the thermodynamic limit at all temperatures studied.

The results for $E/N$ at all values of $T$ that were studied are plotted in Fig.~\ref{fig:CaoBerneEperNVsT}. These values are for the $N=400$ as this satisfies the thermodynamic limit. At larger $T$, $E/N$ becomes linear, as would be expected from a classical system. At lower values of $T$ the system moves away from this classical behaviour as the curve begins to flatten and the slope decreases. This behaviour is expected as $T$ becomes smaller causing the thermal de Broglie wavelength to grow and quantum effects begin to dominate the system's behaviour. 

\begin{figure}[t]
\centering
\includegraphics[width=0.45\textwidth]{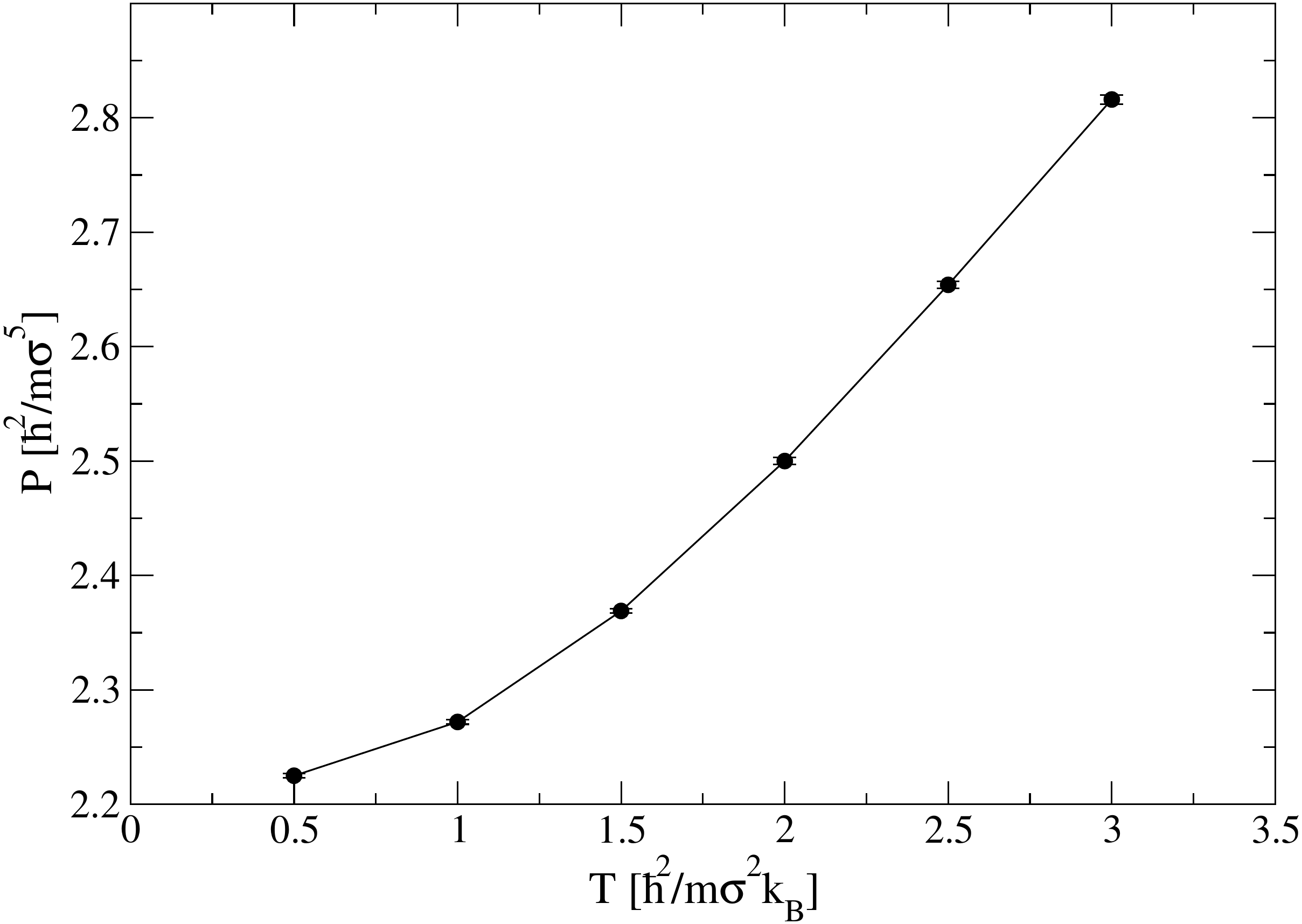} 
\caption{Pressure of system of hard-sphere particles at the thermodynamic limit for density $n \sigma^3=0.2063$. Similar to results for $E/N$, at higher temperatures $P$ is linear in $T$, while this again stops being the case as $T$ is decreased.}
\label{fig:CaoBernePVsT}
\end{figure}     

Fig.~\ref{fig:CaoBerne300PartCv} shows the specific heat of the system which was calculated via the numerical derivative of $E/N$ at many values of $T$ for $N=300$, which was still verified to be within the thermodynamic limit for $T>0.5$. At high $T$ we again observe classical behaviour as $c_v$ approaches $1.5$, given by the equipartition theorem. At low $T$ we can again observe behaviour that deviates from classical expectations as $c_v$ begins to decrease. By the third law of thermodynamics we expect $c_v$ to go to 0 as $T$ goes to 0, however, $c_v$ was calculated at values in the $T<0.5$ range and a rapid increase in $c_v$ was observed as the temperature was decreased in this lower range. Given the third law of thermodynamics, these results appear to be unphysical and could be the result of finite-size effects. This is considered a strong possibility due to the fact that an increase in $c_v$ as the temperature was lowered was also seen for the non-interacting case in Fig.~\ref{fig:FreeGasSpecificHeat_2063rho}, which was also shown to be the result of finite-size effects.

The pressure results plotted against temperature are shown in Fig.~\ref{fig:CaoBernePVsT}. These results are also for the $N=400$ system. In a similar manner to the $E/N$ results, at high $T$ the pressure is shows a linear relation to the temperature, as expected from a classically behaving system. As temperature is lowered we again see the linear relation be begin to flatten and move away from the classical behaviour.  
    
\

\section{Summary and Conclusions}
\label{sec: conc}

In summary, we performed Path Integral Monte Carlo simulations for systems of distinguishable particles that interact via hard-sphere and hard-cavity potentials. To begin, we studied a system of two hard-sphere particles trapped inside a hard-cavity. We analytically calculated the energy of the system at varying temperatures by solving the Schr\"odinger equation and finding the thermodynamic average using the Boltzmann distribution. We calculated the energy of the same system using the PIMC method with two distinct approximations to the thermal density matrix, the Image Approximation, and the CB thermal density matrix. For all temperatures studied, $T(\hbar^2/m\sigma^2k_B)^{-1}=0.5,1.0,1.6,2.0,2.5$, we found that convergence of the PIMC energy to the analytic energy in number of time slices for both density matrices was approximately the same. We then studied a system of $N$ hard-sphere particles placed under periodic boundary conditions. We performed calculations of the energy per particle, pressure and specific heat of the system for a range of temperatures. We established when the thermodynamic limit of the system was reached and the finite-size effects caused by the PBC had been eliminated. A range of $N\approx300-400$ was found to be sufficient. We found that $E/N$, $P$, and $c_v$ approached classical behaviour in the upper limit of the temperature range we studied and deviated from this behaviour at lower $T$. These results constitute non-perturbative microscopic benchmarks
for strongly interacting quantum boltzmannons and can guide further theoretical work as well as comparison with experiment.

\

\begin{acknowledgements}
The authors would like to thank J.\ Carlson  and D.\ T.\ Son for enlightening discussions. This work was supported in part by the Natural Sciences and Engineering Research Council (NSERC) of Canada, the Canada Foundation for Innovation (CFI), and the Early Researcher Award (ERA) program of the Ontario Ministry of Research, Innovation and Science. Computational resources were provided by SHARCNET and NERSC.
\end{acknowledgements}

\end{document}